\documentclass[sigplan,nonacm]{acmart}
\usepackage{algorithm,algpseudocode}
\usepackage{amsmath}

\usepackage{tikz}

\makeatletter
\newenvironment{btHighlight}[1][]
{\begingroup\tikzset{bt@Highlight@par/.style={#1}}\begin{lrBebox}{\@tempboxa}}
{\end{lrbox}\bt@HL@box[bt@Highlight@par]{\@tempboxa}\endgroup}

\newcommand\btHL[1][]{%
  \begin{btHighlight}[#1]\bgroup\aftergroup\bt@HL@endenv%
}
\def\bt@HL@endenv{%
  \end{btHighlight}%
  \egroup
}
\newcommand{\bt@HL@box}[2][]{%
  \tikz[#1]{%
    \pgfpathrectangle{\pgfpoint{0pt}{0pt}}{\pgfpoint{\wd #2}{\ht #2}}%
    \pgfusepath{use as bounding box}%
    \node[anchor=base west, fill=orange!30,outer sep=0pt,inner xsep=0.2em, inner ysep=0.1em,  #1]{\usebox{#2}};
  }%
}
\makeatother

\AtBeginDocument{%
  \providecommand\BibTeX{{%
    \normalfont B\kern-0.5em{\scshape i\kern-0.25em b}\kern-0.8em\TeX}}}

\begin{document}
\newcommand{\Path}{\mathcal{X}}
\newcommand{\Domain}{\mathbf{\Omega}}

\newcommand{\Lvar}{\mathbf{L}}
\newcommand{\Ls}{\Lvar_s}
\newcommand{\Las}{\tilde{\Lvar}_s}
\newcommand{\Lad}{\tilde{\Lvar}_H}
\newcommand{\Lp}{\Lvar_+}
\newcommand{\Ln}{\Lvar_-}
\newcommand{\Ld}{\Lvar_H}
\newcommand{\Ldd}{\Lvar_d}

\newcommand{\Ldelta}{\Lvar_\Delta}
\newcommand{\Ldeltai}{\Lvar_{\Delta i}}
\newcommand{\Ldeltaj}{\Lvar_{\Delta j}}
\newcommand{\Var}{\mathbb{V} }
\newcommand{\Lhp}{\hat{\Lvar}_+}
\newcommand{\Lhn}{\hat{\Lvar}_-}

\newcommand{\Mask}{\mathbb{M}}
\newcommand{\MaskInv}{\Mask^{-1}}

\algnewcommand\bsdf{\mathsf{bsdf}\space}
\algnewcommand\True{\textbf{true}\space}
\algnewcommand\False{\textbf{false}\space}

\newcommand{\red}[1]{\textcolor{red}{#1}}
\definecolor{ForestGreen}{RGB}{34,139,34}

\newcommand{\green}[1]{\textcolor{ForestGreen}{#1}}

\begin{teaserfigure}
	\centering
	\setlength{\tabcolsep}{1pt}
	\begin{tabular}{c}
		\includegraphics[width=\textwidth]{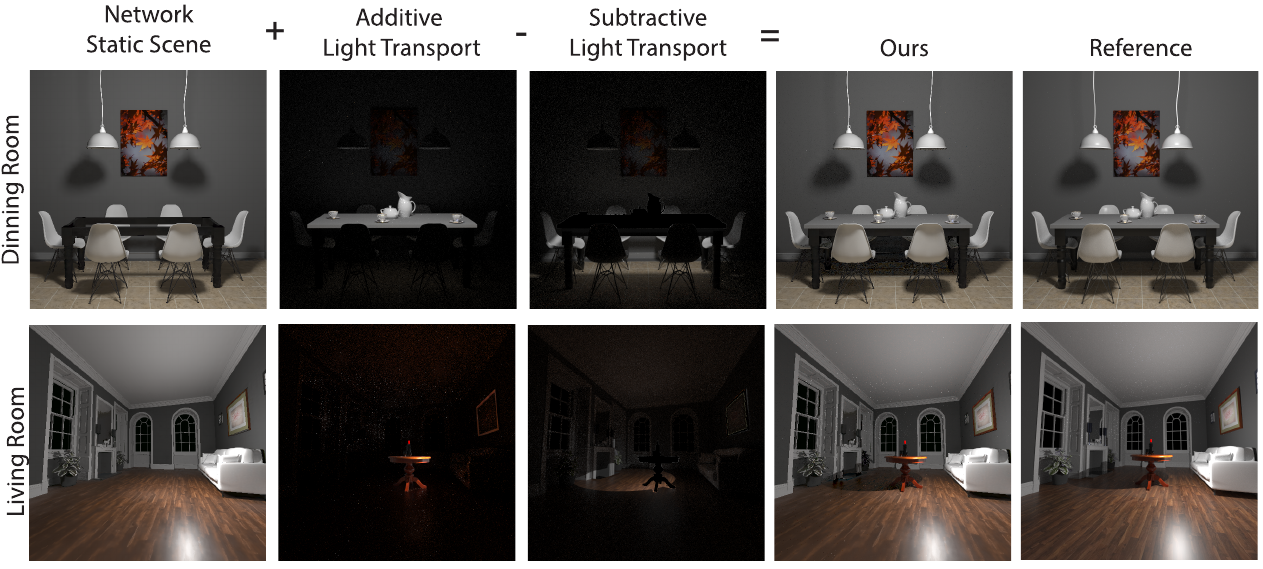} \\
	\end{tabular}
	\caption{
Inserting arbitrary dynamic objects into static scenes. Our method can render it efficiently by taking a static scene and then adding the light transport difference: $\Ldelta = \Lp$ - $\Ln$. By using a pretrained network to represent the light transport in the static scene, we can quickly "render" it without path-tracing and any noise. The difference light transport, $\Ldelta$ has much less noise than rendering from scratch. }
	\label{fig:teaser}
\end{teaserfigure}

\title{Hybrid Rendering for Dynamic Scenes}

\author{Alexandr Kuznetsov}
\email{alexandr.kuznetsov@intel.com}
\affiliation{\institution{Intel Corporation} \country{USA}}

\author{Stavros Diolatzis}
\email{stavros.diolatzis@intel.com}
\affiliation{\institution{Intel Corporation} \country{France}}

\author{Anton Sochenov}
\email{anton.sochenov@intel.com}
\affiliation{\institution{Intel Corporation} \country{USA}}

\author{Anton Kaplanyan }
\email{anton.kaplanyan@intel.com}
\affiliation{\institution{Intel Corporation} \country{USA}}


\begin{abstract}
Despite significant advances in algorithms and hardware, global illumination continues to be a challenge in the real-time domain. Time constraints often force developers to either compromise on the quality of global illumination or disregard it altogether.
We take advantage of a common setup in modern games: having a set of a level, which is a static scene with dynamic characters and lighting.
We introduce a novel method for efficiently and accurately rendering global illumination in dynamic scenes. Our hybrid technique leverages precomputation and neural networks to capture the light transport of a static scene.
Then, we introduce a method to compute the difference between the current scene and the static scene, which we already precomputed.  By handling the bulk of the light transport through precomputation, our method only requires the rendering of a minimal difference, reducing the noise and increasing the quality.
\end{abstract}



\keywords{Neural Rendering, Path-Tracing, Hybrid Rendering, Global Illumination}

\maketitle

\section{Introduction}
Rendering photo-realistic images is a fundamental problem in computer graphics. 
With advances in hardware, physically accurate path tracing became possible. This allows physically accurate light simulations. 
However, full path tracing is an expensive endeavor, especially in real-time applications such as games. With the current hardware, even in simple scenes, the amount of rays per pixel is quite limited. As a result, heavy denoising needs to be applied which significantly blurs and degrades the quality of an image.
 An alternative option is to precompute global illumination via deep learning for example. This works extremely well in static scenes. However, it fails when anything changes in the scene. This is because of the curse of dimensionality. Instead of learning a 4D function (2 for space and 2 for directions) for a static scene, for a dynamic scene, you need to learn a 7D+ function, requiring enormous datasets and long training.  Besides, such systems cannot display novel objects, which severely limits their use cases. Therefore,  pure precomputation for dynamic scenes is not practical.
In this paper, we combine the benefits of using precomputations for a static portion of a scene and using path tracing for the dynamic portion of the light transport. 

One of the major applications of our method is games. Real-time global illumination is challenging and it slowly becoming a reality due to advances in hardware raytracing. 
Games are usually made of levels, which are mostly static with movable objects and characters. 
In those applications, a scene is rendered in every frame. However, the light transport in the scene doesn't change much from frame to frame. We want to take advantage of this property in order to accelerate the rendering of global illumination. 
To do that, we separate the scene into static and dynamic parts. Assuming we can render the static parts of the scene quickly with the help of a precomputation, the task is to render the changes caused by the dynamic part of the scene. In this paper, we will look at how we can render changes in the light transport caused by dynamic objects and lights.

The goal of the paper is to introduce the idea of the hybrid rendering. THe faster rendering more realistic computer graphics in real-time domains such as games.
The key contributions of our paper are as follows:
\begin{itemize}
    \item We introduce the idea of hybrid rendering of static scenes with \emph{arbitrary} dynamic objects and dynamic lights. By separating a scene into precomputed static and sparse dynamic light transport, we can more efficiently render both. 
    \item Our method supports the addition of arbitrary dynamic objects.
    \item We introduce a mathematical justification and show that the method is conditionally unbiased.

    \item We also make the connection to the Primary Space Sampling.
    
\end{itemize}

Our method combines the benefits of precomputation and adaptive path-tracing, to render efficiently render global illumination in real-time. 
\section{Related Work}

In this section, we review related work from path tracing, precomputation techniques, neural rendering \& denoising and finally adaptive approaches. Each one of these fields is active and vast so we review methods closest to our approach, while referring to surveys whenever possible for an extensive overview.

\subsection{Path Tracing}

Monte Carlo estimators in the context of path tracing have been used extensively to create realistic-looking images of synthetic scenes and a lot of research has gone into improving their sampling quality to generate these images faster. These images are plagued with noise, inherent in Monte Carlo estimators and different variance reduction techniques, such as importance sampling, control variates and sample reuse have been proposed to get cleaner images with less samples. 

In the past the evaluation of visibility between two surfaces was too costly for interactive applications leading to innovative work like~\cite{dachsbacher2007implicit} which avoided this computation by implicitly defining visibility through antiradiance. Our method is inspired by this concept but we aim to use our formulation only for the parts of a scene that have changed given a precomputed image/radiance field.

The emergence of path tracing hardware \cite{keller2015path} opened the door to real-time path tracing, a long-standing goal of computer graphics. Despite the recent advances in hardware, the sampling budget in interactive applications is still restricted to a few samples per pixel, compared to the hundreds or thousands used in offline rendering. With this restriction in mind, ReSTIR and variants~\cite{bitterli2020spatiotemporal, ouyang2021restir, lin2022generalized} propose to reuse samples either temporally or spatially, increasing the effective sample count in cases of coherent content between frames and pixels. Reuse is less useful for hard-to-sample effects, such as caustics and specular-diffuse-specular paths, which need hundreds of samples to even appear and as a result, they are either treated explicitly or avoided. Our method is orthogonal to ReSTIR and can be easily integrated to work together with ReSTIR.  

\subsection{Precomputation Techniques}

Related to our method, there is a long history of precomputation techniques that use information from a one-time preprocessing step to improve the quality of rendering during runtime. Ward et al.~\cite{ward1988ray} precomputes the outgoing radiance within a scene and stores it explicitly in a spatial cache. During rendering an interpolation scheme is used, ignoring view dependency, to infer the outgoing radiance at any point in the scene. This was improved upon by ~\cite{krivanek2005radiance} to take into account the viewing direction encoded by spherical harmonics while illumination gradients improved the interpolation. In~\cite{zhou2005precomputed} the authors suggest the precomputation of shadow fields and use them to efficiently compute soft shadows for dynamic scenes. Precomputation using low rank operators and proxy shapes was introduced in~\cite{loos2011modular} and improved upon in~\cite{loos2012delta} which has the advantage of being a scene-independent precomputation technique but suffers in terms of quality compared to the ground truth.

More recently, many pre-baking systems~\cite{seyb20uberbake, silvennoinen2021moving} have been developed using multiple components and data structures that have been engineered to enable dynamic changes to some aspects of the precomputed radiance transfer. Our methods explore how real-time path tracing can augment images of static components in a scene with dynamic content. The static component can also be precomputed as in this previous work.

\subsection{Neural Rendering \& Denoising}

Neural networks have been used to render images in the context of synthetic scenes, novel view synthesis and generative models in the rapidly growing field of neural rendering. An overview of the field can be found in~\cite{tewari2022advances}. We focus on the use of neural networks to replace, augment or speed up rendering of synthetic scenes.

A first attempt at using neural networks to regress the radiance of a given scene was proposed in ~\cite{ren2013global}. Here a Multi-Layer Perceptron (MLP) is trained to infer the outgoing radiance for any point in the scene given the world position, viewing direction, BSDF parameters etc. In a similar spirit, the work by Granskog et al.~\cite{granskog2020compositional} renders a scene from different viewpoints and uses them to create a compositional neural scene representation to shade easy-to-compute G-Buffers. An object-oriented approach proposed by~\cite{zheng2023nelt} composed a scene by combining neural networks that represent each object's impact in light transport.

Operating in image space~\cite{Nalbach2017b} uses a convolutional neural network (CNN) within a deferred shading step to speed up the rendering of effects such as ambient occlusion, depth of field, etc. CNNs have also been shown to be efficient denoisers leading to clean images with much less samples.~\cite{Bako17} uses two deep networks, one for the diffuse and one for the specular component, which take G-buffers as input and are trained to reduce the noise of an input image. Gharbi et al.~\cite{gharbi2019sample} use a denoising network that operates on the input samples of the rendering, instead of the collapsed image. This leads to better performance since the network has more information about the underlying effects of the image. \cite{isik2021anf} proposes the use of a lightweight network to enable denoising at interactive rates while being temporally stable. Denoising can also be combined with supersampling~\cite{thomas2022temporally} enabling rendering in lower sample counts \emph{and} resolution. The path-tracing component of our method can benefit from a denoising/supersampling step to improve the quality of the final image.   

MLPs which operate in world space, similar to~\cite{ren2013global}, have recently gained popularity due to their inherent view consistency and ability to implicitly represent the radiance fields of a given scene from sparse training data. In~\cite{NeuralCache} a small network is used as a radiance cache which is constantly trained and updated on the fly. This cache is used to guide paths and terminate them early. This approach enables the use of a predicting neural network in a fully dynamic scene but comes with the trade-off of allocating a compute bandwidth to constantly train and update the network. To pretrain an MLP and learn the radiance field for a variable scene,~\cite{diolatzis2022active} encode the predetermined variability of a scene into a conditioning vector. While this shows promising quality for traditionally hard-to-render effects the variability is limited to a few predetermined factors. We aim to alleviate this issue of pretrained neural radiance fields by efficiently combining their outputs with path tracing.  

\subsection{Adaptive Methods}

Our method is most closely related to adaptive methods, which guide sample placement during path tracing to reduce the noise in the final image. We specifically focus on adaptive methods that combine path tracing with a neural component and refer to~\cite{zwicker2015recent} for an overview of more traditional approaches.

In~\cite{kuznetsov2018deep} the authors propose a denoising-aware sample placement approach. Two CNNs are trained end to end, one learns to create a sampling map which is used to render the noisy image, and another denoises that image. This leads to increased sample placement in areas where the denoiser needs more information. This also has been shown to work effectively in the temporal domain~\cite{hasselgren2020neural} by using information from temporally reprojected previous frames. Recently analytic distributions~\cite{salehi2022deep} have been proposed to guide the sample placement before a denoising step from denoising performance. Our approach draws inspiration from the adaptive nature of these methods but focuses on the scenario of augmenting a precomputed image/representation with new dynamic content. Parallels can also be drawn between our method and~\cite{rousselle2016image} which formulates the problem as image-based control variates. They use a pre-rendered image with a computed difference given changed materials. However, their method requires a fixed camera, making it impractical in real-world applications. Our method utilizes the recent advances in neural networks to enable a moving camera and adds support for dynamic objects and lights, which opens up a whole new realm of possibilities. Our adaptive sampling exploits the sparsity of the delta light transport to improve the quality even further.

\section{Method}

Our method has two major components: rendering dynamic light transport and rendering static light transport. The basic idea of our method is we can quickly render the light transport of the static scene and then adaptively render the light transport caused by dynamic objects. 
First, we will look into how we can incorporate the rendering of dynamic objects. Then we will take a look at how we can render static scenes. 

\subsection{Rendering Dynamic Light transport}
In order to render a scene with dynamic objects, we want to render the static scene, plus the light  transport delta caused by adding the dynamic objects to the scene:

\begin{figure}[]
\centering
\includegraphics[width=.9\columnwidth]{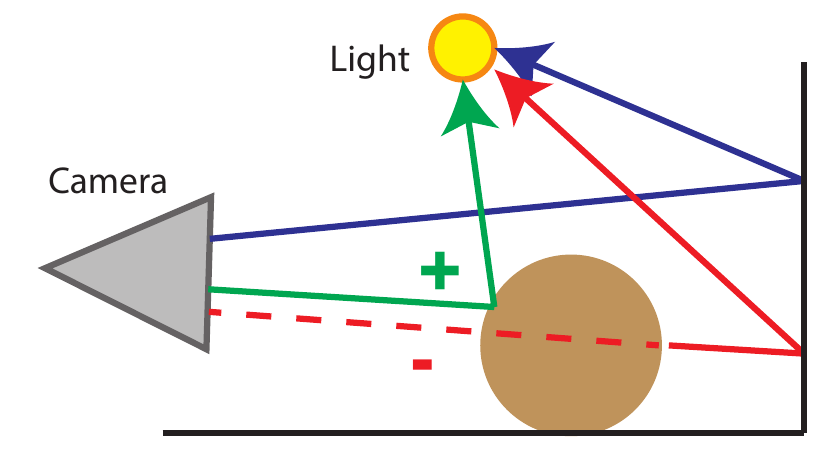}
\caption{Example of different paths during rendering a scene. Here, we have a beige sphere as a dynamic object. The blue path $\in \Domain_o$   doesn't interact with the dynamic object. The green path  $\in \Domain_+$  reflects from the dynamic object. And the red path  $\in \Domain_-$  goes directly through the dynamic object unaffected.  }
\label{fig:scene}
\end{figure}

\begin{equation}
 \begin{split}
    \Ld &= \Ls + (\Lp - \Ln) \\
    \Ld &= \Ls + \Ldelta
     \end{split}
    \label{eq:main}
\end{equation}
where $\Ld$ is the light transport of the scene with the dynamic objects, $\Ls$ is the light transport of the static scene (without the dynamic objects), $\Lp$ is the additive light transport caused by adding the dynamic objects, and $\Ln$ is the subtractive light transport caused by adding the dynamic objects. See Figure \ref{fig:scene} for details. Computing $\Ls$ can be achieved through classical path-tracing or through precomputation as will be discussed in the next section. The crux is how can we compute $\Lp$ and $\Ln$.

\emph{Additive Light Transport. }  Here, we will discuss how we compute $\Lp$. This is an additional radiance contribution caused by adding the dynamic objects. For example, by adding a green sphere, we will get green inter-reflection on the floor. We can define it as the light contribution caused by adding the dynamic objects. So, we start accumulating the radiance only after the path is reflected by a dynamic object.  
Please refer to Algorithm \ref{alg:poslight} for the implementation details. 
The algorithm can easily be extended to support Multiple Importance Sampling.

\newcommand{\HiLi}[1]{#1}
\begin{algorithm}[]
\caption{Calculating $\Lp$}
\label{alg:poslight}
  \begin{algorithmic}[1]
\State $\Lp \gets 0$  \Comment{Additive Radiance}
\State $T \gets 1$  \Comment{Transmittance }
\State  \HiLi{$h \gets \False$} \Comment{Did we hit a dynamic object yet?}
\State $r \gets \textproc{InitialRay}{()}$
\While {true}
\State $\bsdf \gets  \textproc{RayIntersect}{(r, S \cup D)}$\Comment{ Scene with Dynamic}
\State \HiLi{$h \gets h | \textproc{IsDynamic}{(\bsdf)}$}
\If {h} 
\State \HiLi{ $\Lp \gets \Lp + T \cdot \textproc{Emmiter}{(r,\bsdf)}$}
\EndIf
\State $r,T \gets \textproc{SampleBSDF}_+{(\bsdf, r, T)}$
\State $T \gets \textproc{RussianRoulette}{(T)}$
\If{T=0}
\State \textbf{break}
\EndIf
\EndWhile
\State $\Return \: \Lp$
  \end{algorithmic}
\end{algorithm}

\begin{algorithm}[]
\caption{Calculating $\Ln$}
\label{alg:neglight}
  \begin{algorithmic}[1]
\State $\Ln \gets 0$  \Comment{Subtractive Radiance}
\State $T \gets 1$  \Comment{Transmittance }
\State  \HiLi{$h \gets \False$} \Comment{Did we hit a dynamic object yet?}
\State $r \gets \textproc{InitialRay}{()}$
\While {true}
\State $\bsdf \gets  \textproc{RayIntersect}{(r, S \cup D)}$\Comment{ Scene with Dynamic}
\State \HiLi{$d \gets \textproc{IsDynamic}{(\bsdf)}$}
\State \HiLi{$h \gets h | d$}
\If {d} \Comment{Ignore dynamic objects}
\State \HiLi{ $r \gets \textproc{SkipObject}{(r)}$}
\State \HiLi{\textbf{continue}}
\EndIf
\If {h}
\State \HiLi{$\Ln \gets \Ln + T \cdot \textproc{Emmiter}_-{(r,\bsdf)}$} \Comment{Accum. radiance}
\EndIf
\State $r,T \gets \textproc{SampleBSDF}{(\bsdf, r, T)}$
\State $T \gets \textproc{RussianRoulette}{(T)}$
\If{T=0}
\State \textbf{break}
\EndIf
\EndWhile
\State $\Return \: \Ln$
  \end{algorithmic}
\end{algorithm}

\emph{Subtractive Light Transport.}  Here, we will discuss how we compute $\Ln$. This is a subtractive radiance contribution caused by adding dynamic objects. For example, adding a cube in front of the light will cause a shadow that wasn't present in the static scene.
The subtractive contribution or $\Ln$ is just the accumulation of all the light paths in the static rendering, which would go through the dynamic objects if they were present. So, in order to calculate $\Ln$, we just need to re-render the static scene and accumulate only the contributions after the ray goes through a dynamic object. Here, we treat the dynamic objects as totally transparent. 
Please, refer to the Algorithm \ref{alg:neglight} to see how to compute  $\Ln$. Similarly, this algorithm is trivially extended to MIS.

\subsection{Mathematical Justification}

Here we will prove that our definitions the Equation \ref{eq:main} hold true with the given definitions. Classically, the rendering equation is written like this:
 \begin{equation}
  \begin{split}
\Lvar(x, \omega_o) = \Lvar_e(x, \omega_o) + \int f(x, \omega_i, \omega_o) \cdot G(x, x') \, d\omega_i
    \label{eq:rendering} 
 \end{split}
 \end{equation}
 where $\Lvar_e$ is the emission radiance on the surface, $f()$ is the cosine weighted BSDF term and $G()$ is the geometric visibility term.  This equation can be rewritten as an integral over all possible paths in the scene between the camera and all the emitters:
 \begin{equation}
  \begin{split}
\Lvar(x, \omega_o) =  \int_{\Domain(x, \omega_o)} T(\Path) \Lvar_e(\mathrm{last}(\Path)) \, d\Path = I(\Domain(x, \omega_o))
    \label{eq:rendering} 
 \end{split}
 \end{equation}
where $\Domain(x, \omega_o)$ is a set of possible paths $\Path$ in the scene which start at $x$ and direction $\omega_o$ and end at an emmiter, and $T()$ is the transmittance of the path. As a shortcut, from now on we will refer to $\Domain(x, \omega_o)$ as simply $\Domain$. 

Now let's define the following path domains as a subset of all possible paths in a space: $\Domain_o$ - a path that doesn't go through any dynamic object (blue in Fig. \ref{fig:scene}), $\Domain_-$ - a path that ignores dynamic objects (red), and $\Domain_+$ - a path that interacts with every dynamic object it encounters (green). 
See figure \ref{fig:scene} for the illustration. Those are disjoint sets, s.t.: 
 \begin{equation}
  \begin{split}
  \Domain_o\cup\Domain_-\cup\Domain_+ = \Omega
  \end{split}
 \end{equation}

Then, we can rewrite $\Ls$ and $\Ld$ as follows:
 \begin{equation}
  \begin{split}
\Ls = I(\Domain_o) + I(\Domain_-) \\
\Ldd = I(\Domain_o) + I(\Domain_+) 
 \end{split}
 \end{equation}

 By combining the equations we get:

  \begin{equation}
  \begin{split}
\Ldd &= I(\Domain_o) +  (I(\Domain_-)   - I(\Domain_-))  + I(\Domain_+) \\
 &= (I(\Domain_o) +  I(\Domain_-))   - I(\Domain_-)  + I(\Domain_+) \\
&= \Ls  - I(\Domain_-)  + I(\Domain_+) \\
&=\Ls  - \Ln + \Lp = \Ls + \Ldelta
 \end{split}
 \end{equation}

 As a result, we can see that our method is unbiased as every integrator is unbiased. Later, we will replace the static scene renderer  $I(\Domain_o)$, with a neural network, which is biased.

\subsection{Primary Sample Space} 

Primary Sample Space is the idea of operating on a hypercube of random numbers. It was first popularized by Kelemen et al. \cite{kelemen2002simple} to simplify the Metropolis Light Transport algorithm. We found utilizing primary sample space also simplifies  the integration of our algorithm into existing engines. We used this method, to integrate our method with the real-time path tracer Falcor. Later, we will show that those two approaches are equivalent. PSS duplicates the computation of the paths that haven't diverged. However, we didn't find a severe negative impact on performance. 

The algorithm is described in \ref{alg:deltalight}.
Instead of tracing the ray until we hit a dynamic object and splitting the ray into two, we render the same scene twice. First, we render the static scene and then we render the dynamic scene. If we do that naively,  the noise will be not correlated and we get the different image which is noisy. However, if we use primary sample space and reuse the random numbers for the both renders, the noise pattern will be correlated. This is because it will diverge only if the ray hits a dynamic object. As a result, the different images will have much less noise compared to individual renderings. 

Using primary sample space algorithm \ref{alg:deltalight} is equivalent to calculating the difference we get from \ref{alg:poslight} and \ref{alg:neglight} algorithms. This is because until we hit a dynamic object, we trace the same paths. If we use the same random numbers in \ref{alg:poslight} and \ref{alg:neglight}, the path will be exactly the same. Therefore, it will produce the same result. 

\subsection{Static Scene Rendering} 
Although we can render simply $\Ls$, this will defeat the purpose of the method as rendering $\Ld$ is not much more expensive. Therefore, we want to precompute the light transport for the static scene. Because the scene is static, this is a relatively simple thing to do. We need to learn just a 5D function $\Las(x,\omega)$ - 3 dimensions for the position of the ray and 2 for direction. This task can be approached in a number of ways, but we chose to utilize a hash grid with an MLP. Inspired by  \cite{NeuralCache}, given a 3D location of the intersection, we look up a corresponding latent vector in the hierarchical hash grid. Then, we concatenate the latent vector with camera direction and surface normal vectors and pass it to the MLP.  The MLP consists of 7 hidden layers with 64 channels in each. It outputs the final radiance information. As the network is small and operates in screen space (unlike volumetric NeRF-like methods), the runtime is negligible.

In order to train the network, we first must generate a dataset of the static scene. This is done by originating the rays at random inside the scene going to random directions. We used  $1 \times 10^9$ ray samples to train our network.  Unlike \cite{NeuralCache}, we pretrain the network offline for 2 hours and only do inference during the runtime. That way, we don't need to do ray-tracing and training during the runtime, saving us the compute at the expense of bias. 

\subsection{Adaptive Sampling} 
Adaptive Sampling is a crucial part of our method. $\Ldelta$, the difference in light transport is very sparse. The majority of the difference is located near a dynamic object or a dynamic light. Therefore, we can use adaptive sampling to focus on those regions. Because those regions are typically small compared to the overall image size, we can spend much more samples in those regions. Because in the rest of the image, $\Ldelta$ is close to zero, allocating fewer samples there doesn't contribute to a higher error.

There are different strategies for adaptive sampling. One way is to make an initial pass at a lower resolution or sampling count in order to estimate the sampling map as it was done in \cite{kuznetsov2018deep}. The other way is to exploit the temporal information from the previous frames to estimate the sampling map for the current frame as was done in \cite{hasselgren2020neural}. The advantage of this approach is that there are no extra samples you need to render.

There are different ways to estimate the sampling map from the SURE-based estimator to a neural network. However, we opted for a simpler approach: the samples are distributed proportional to the variance and intensity of the samples. We found this approach is really fast and generates sufficiently good sampling maps. 
 \begin{equation}
s_i = S\cdot \frac{\sigma(\Ldeltai) + |\Ldeltai|}{1/N\sum_j \sigma(\Ldeltaj) + |\Ldeltaj|} 
    \label{eq:adaptive} 
 \end{equation}
 where $s$ is the sample count for the  pixel $i$,  $S$ is the target sample count,  and $N$ is the total number of pixels in the image. Because the estimates of the variance and intensity are noisy, we apply a simple Gaussian filter with a kernel of size 5, which removes the noise. Then we normalize the sampling map to the desired sampling count and apply a dither operation to discretize the sampling map.:
 \begin{equation}
\hat{s}_i = \text{round}(s_i + U)
    \label{eq:adaptive2} 
 \end{equation}
where $U$ is the random value between -.5 and 0.5

With this direct approach, two problems arise: bias and exploration. Some regions of the image are of very low intensity and variance, so they get a total of 0 samples per pixel across the whole image. This introduces bias because we assume that there are no differences, while there might be.  And second of all, because we use previous frames to predict the sampling maps, it creates a positive feedback loop where we don't render anything, so the variance is zero, and so on.  To mitigate this issue, we set the minimum number of samples to be 1 per 4 pixels. That way we always have an estimate of variance and intensity anywhere in the image. This solves the sampling map problem, but not the bias. Inspired by Russian Roulette, we propose to divide the intensity by the pre-quantized sampling map. 
 \begin{equation}
\hat{\Ldelta} =  \frac{\Ldelta}{\min(1, s)}
    \label{eq:boost}
 \end{equation}
That way we compensate for the pixels we didn't render.

\begin{figure}[]
\centering
\includegraphics[width=.5\columnwidth]{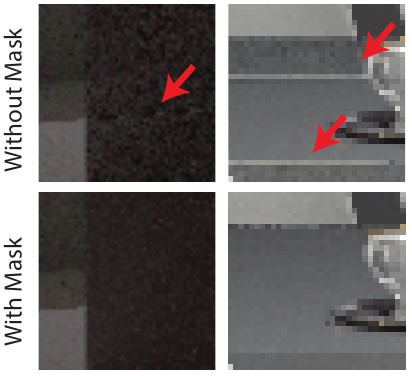}
\caption{Example of hybrid rendering with and without the mask. Without the mask, we get visible artifacts at the edges due to the difference between the real base image and the learned-based image. The background error is leaking through the dynamic objects. Due to the nature of artifacts, it's hard for a denoiser to get rid of them.   }
\label{fig:artifacts}
\end{figure}

.

See the algorithm \ref{alg:deltalight} for details.

\begin{algorithm}[]
\caption{Calculating difference using PSS}
\label{alg:deltalight}
  \begin{algorithmic}[1]
\State $\Lp \gets 0$  \Comment{Radiance of the new scene}
\State $\Ln \gets 0$  \Comment{Radiance of the static scene}
\State $T_+ \gets 1$  \Comment{Transmittance of new}
\State $T_+ \gets 1$  \Comment{Transmittance  of static}
\State $u_0,u_1, u_2 \ldots \gets \textproc{InitRandomSequence}{()}$  \Comment{List of random numbers }
\State $r_+ \gets  \textproc{InitialRay}{()}$
\State $r_- \gets  \textproc{InitialRay}{()}$
\For{$i \gets 0$ to $\infty$}  
\State $r_+,T_+ \gets \textproc{IntersectAndSample}{(u_i, \bsdf_+, r_+, T_+)}$
\State $r_-,T_- \gets \textproc{IntersectAndSample}{(u_i, \bsdf_-, r_-, T_+)}$
\State \HiLi{ $\Lp \gets \Lp + T_+ \cdot \textproc{Emmiter}_+{(r_+,\bsdf_+)}$}
\State \HiLi{ $\Ln \gets \Ln + T_- \cdot \textproc{Emmiter}_-{(r_-,\bsdf_-)}$}
\State $T_+, T_- \gets \textproc{RussianRoulette}{(T_+, T_-)}$
\If{T=0}
\State \textbf{break}
\EndIf
\EndFor
\State $\Return \: \Lp - \Ln$
  \end{algorithmic}
\end{algorithm}

\subsection{Dynamic Lights} 
In games and other scenarios, it's common for the lighting to change. Our method can also support changing illumination. When a  light moves, it doesn't affect the whole scene equally. Due to the inverse square law, the moving light affects surfaces closest to the light source the most. For example, if you move a light near a wall, the wall's brightness changes dramatically, but the overall illumination of a room doesn't change. Our algorithm algorithm can support dynamic lights combined with dynamic objects. Please, see Fig. \ref{fig:comp_lights} for examples. This requires a simple change in the light evaluation function. When evaluating the light contribution for $\Lp$ and $\Ln$, we need to use the corresponding state of the light.

\begin{figure}[]
\centering
\includegraphics[width=1\linewidth]{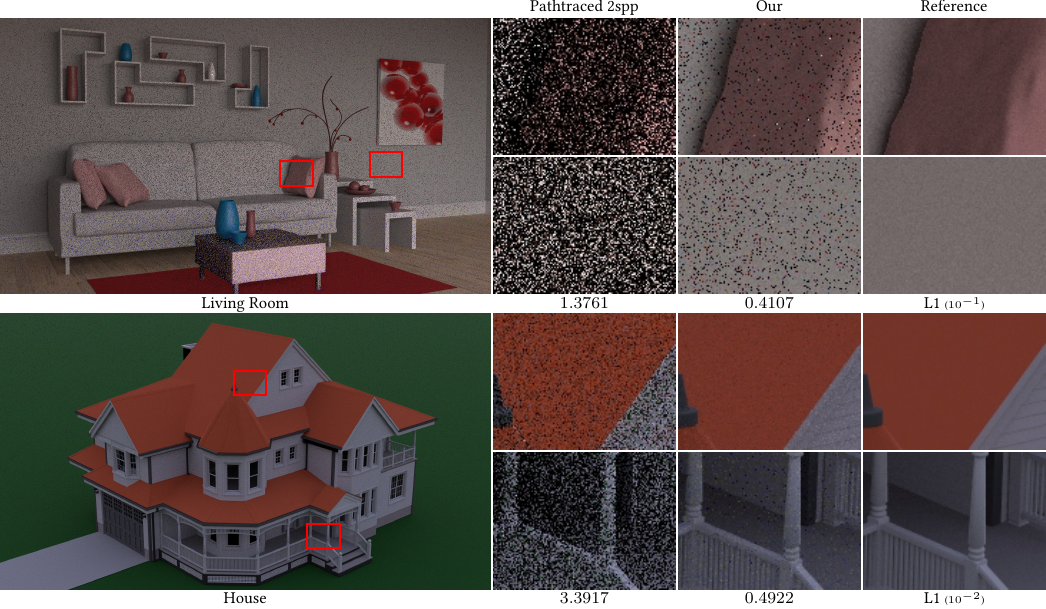}
\vspace{-12pt}
\caption{
Scenes illuminated by changing environment lighting. In path tracing, we need to render everything from scratch. In our method, we only render the difference, leading to less noise.
}
\vspace{-12pt}
\label{fig:comp_env}
\end{figure}

{Environment Maps}. Often a scene is illuminated by an environment map.  Often that environment map changes from frame to frame. For example, clouds move through the sky or the sun moves. It would be a complete waste to start from scratch and completely re-render the scene. Instead, we want to use an existing rendering and modify it to reflect a new environment map. See Fig \ref{fig:comp_env} for the example. To improve the quality further, we employ the standard important sampling.
Unlike the traditional environment maps, $E_\Delta$ might have negative values.
Therefore, when we import sample the environment map, we should sample proportionally to $|E_\Delta|$.

\begin{figure*}[]
\centering
\includegraphics[width=.9\linewidth]{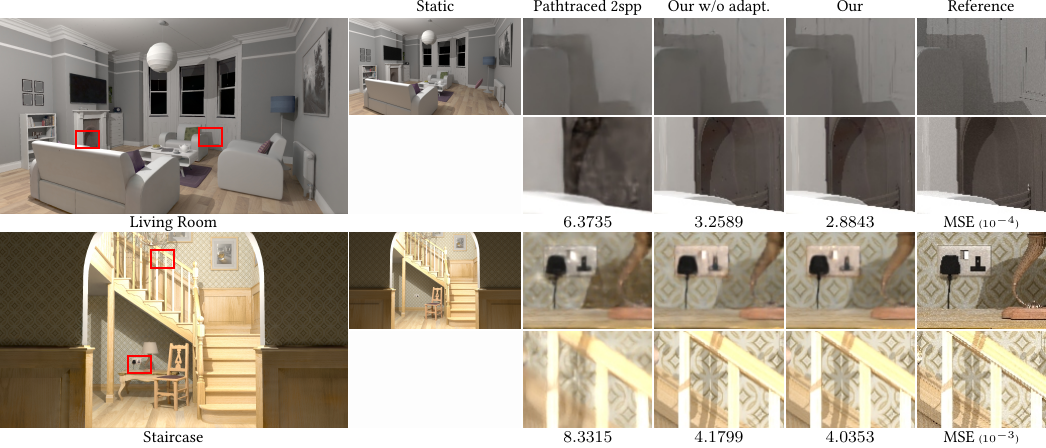}
\vspace{-12pt}
\caption{
Example of our method with a denoiser applied. The 1st column is path-traced images with 2 spp.  The 2nd column is our method with just one 1spp. The next to last column is our method with adaptive sampling. Even without the adaptive sampling, our method is much better compared to the path-tracer. Our method lowers the noise level in the input to the denoiser, which improves the quality of the denoiser. 
}
\label{fig:comp_denoise}
\end{figure*}

\subsection{Masked Rendering} 
\label{sec:masked}
Although the equation \ref{eq:main} is mathematically correct and unbiased, it assumes that $\Ls$ is also unbiased. As we have seen from the previous section, this is might not the case. Because of the biases in $\Las$, the equation \ref{eq:main} is no longer valid. This causes artifacts as shown in Figure \ref{fig:artifacts}.  To avoid this, we propose to mask out the direct contribution of the dynamic objects. Let $\Mask$ be the mask that excludes all dynamic objects in the first bounce.  Then:

\begin{equation}
    \Lad \approx \Mask \times (\Las - \Ln) + \Lp
    \label{eq:masked}
\end{equation}

From Algorithms \ref{alg:poslight} and \ref{alg:neglight}, we can see that if we hit the dynamic object first, then $\Ld=\Lp$ and $\Ls = \Ln$. So, we can safely ignore the non-zero error $\Las - \Ln$ when we hit a dynamic object first

\subsection{Denoising and Other Improvements} 
Our method can be used with a denoiser as seen in Fig. \ref{fig:comp_denoise}. Our method decreases the noise, which improves the quality of denoised images.

Our method is orthogonal to other techniques used to accelerate the rendering. For example, our method can be easily used with path-guiding methods. Instead of using the SampleBSDF function, we substitute it with a sampling function from a path-guiding algorithm. A small modification that's needed is to use the same path-guiding state across algorithms \ref{alg:poslight} and \ref{alg:neglight}. That way the sampling function would behave identically. A further improvement can be achieved by training path-guiding on the difference $\Ldelta$ instead of overall intensity.

Similarly, it can be combined with ReSTIR \cite{bitterli2020spatiotemporal}. 
ReSTIR improves the sampling of lights. Similar to the path-guiding case,  we just need the same state across algorithms \ref{alg:poslight} and \ref{alg:neglight}. That way samples will stay correlated. 
\section{Results}

In this section, we discuss the performance of our method across different scenes in a variety of scenarios. First, we will discuss how our method performs in adding an object to a scene scenario. Then, we will see how our method helps a denoiser generate better images.
Then, we will see what will happen if we also change the lighting. Then, we will see how we can use our method with changing environment maps.

In figure \ref{fig:comp1} we have a comparison of our method to path tracing. Here, we inserted a completely novel object. It alters the light transport of a scene in a nontrivial way. The leftmost images are images using our method. Here, we used a neural network to compute the base image of the static scene and only 1spp on average to compute the difference. 
Next, we have images of the static scenes. We used the complete light transport of those scenes to train the neural network. In the first column of the comparisons, we have a path-traced image using 2spp. As you can, the results are quite noisy everywhere, which makes it hard to see the details of complex scenes. If we want to use a path-tracer in games, it's clear that 2spp is not enough.
Next, we have our method without adaptive sampling. It uses just 1spp, which is equivalent to 2spp in a path tracer because we need to trace both $\Lp$ and $\Ln$ contributions. Here, the images are still noisy where the dynamic objects are (top row), but they are almost noise-free in the areas that aren't dynamic objects. This is because we used a neural representation for the base image, which is noise-free. So, the noise is only coming from the  $\Lp$ - $\Ln$  difference. Because those samples are correlated, the error is much lower than from path tracing. 
The second to last column is our method with adaptive sampling. In the previous column,  some regions were noise-free, while others were quite noisy. We can redistribute the samples to the regions where they are needed the most. By doing so, we increase the quality without increasing the sampling budget. Our method resembles the reference images even without a denoiser. Please see the supplemental materials for full-sized images. 

We can use a denoiser to improve the result even further. In figure \ref{fig:comp_denoise} we applied the Optix denoiser. As you can see, the path-traced result is quite blurry. For example, there is not enough information to reconstruct the fireplace in the living room scene or the shadow in the staircase scene.  This is because the inputs are quite noisy as we saw in Fig. \ref{fig:comp1}. Our method makes the input images less noisy, which translates to a better reconstruction. Unlike traditional path-tracers, the characteristics of our noise are quite different. It's a sparse pepper and salt whose value can be actually negative. Despite this, the pretrained denoiser handles our method quite well. But we believe that training a denoiser directly on our method would have improved our results even further, 

Not only an object can be changed in a scene. Often the lighting changes as well. In fig. \ref{fig:comp_lights} we have the scenes where not only an object was added, but also the lighting has changed. This is a more challenging scenario because $\Ldelta$ is no longer that sparse. Nevertheless, our method works pretty well in all the scenes.

In Fig. \ref{fig:comp_env} we show the results with changing environment maps. For the path-tracer, we need to integrate the environment map, which requires a lot of samples. Many contributions here are zero because they are coming from behind. Our method works much better because we sample the difference in the environment maps.

Video comparisons. We also include video comparisons in the supplemental materials. Our method is temporally stable. Moreover,  because our method  has a lower noise level, the denoiser can handle the input better, creating reference-like results without flickering artifacts. We implemented our method in Falcor and we achieve real-time performance (30-50fps depending on a scene). The FPS depends on scene complexity, but 1 spp of our method is about the same performance as the  path-tracer with 2spp  because we need to trace for $\Lp$ and $\Ln$. The static scene evaluation is negligible because it's done on the auxiliary  buffers on tensor cores part of the GPU.

\section{Conclusion and Future Work}
In conclusion, we presented a novel method for rendering hybrid scenes: a static scene with {arbitrary} dynamic objects and dynamic lights. We used a neural network to represent the static scene and use then path traced the difference. The samples are correlated, leading to smaller errors and faster convergance time.  Our approach is opening new possibilities in the rendering of global illumination in real-time applications like games, where you have mostly static levels with movable characters.
As we have shown, $\Lp$ and $\Ln$ are really sparse. Therefore we utilize adaptive sampling to smartly allocate samples to reduce the noise.   In future works, we would like to explore the importance sampling of dynamic objects in order to reduce the error.  Also, we used off-the-shelf denoiser, which worked really well. We would like to design a denoiser that can better handle the negative noise.

\bibliographystyle{ACM-Reference-Format}
\bibliography{main}


\begin{thebibliography}{29}


\ifx \showCODEN    \undefined \def \showCODEN     #1{\unskip}     \fi
\ifx \showDOI      \undefined \def \showDOI       #1{#1}\fi
\ifx \showISBNx    \undefined \def \showISBNx     #1{\unskip}     \fi
\ifx \showISBNxiii \undefined \def \showISBNxiii  #1{\unskip}     \fi
\ifx \showISSN     \undefined \def \showISSN      #1{\unskip}     \fi
\ifx \showLCCN     \undefined \def \showLCCN      #1{\unskip}     \fi
\ifx \shownote     \undefined \def \shownote      #1{#1}          \fi
\ifx \showarticletitle \undefined \def \showarticletitle #1{#1}   \fi
\ifx \showURL      \undefined \def \showURL       {\relax}        \fi
\providecommand\bibfield[2]{#2}
\providecommand\bibinfo[2]{#2}
\providecommand\natexlab[1]{#1}
\providecommand\showeprint[2][]{arXiv:#2}

\bibitem[Bako et~al\mbox{.}(2017)]%
        {Bako17}
\bibfield{author}{\bibinfo{person}{Steve Bako}, \bibinfo{person}{Thijs Vogels},
  \bibinfo{person}{Brian McWilliams}, \bibinfo{person}{Mark Meyer},
  \bibinfo{person}{Jan Nov\'ak}, \bibinfo{person}{Alex Harvill},
  \bibinfo{person}{Pradeep Sen}, \bibinfo{person}{Tony DeRose}, {and}
  \bibinfo{person}{Fabrice Rousselle}.} \bibinfo{year}{2017}\natexlab{}.
\newblock \showarticletitle{Kernel-Predicting Convolutional Networks for
  Denoising Monte Carlo Renderings}.
\newblock \bibinfo{journal}{\emph{ACM Transactions on Graphics (TOG)
  (Proceedings of SIGGRAPH 2017)}} \bibinfo{volume}{36}, \bibinfo{number}{4}
  (\bibinfo{year}{2017}).
\newblock


\bibitem[Bitterli et~al\mbox{.}(2020)]%
        {bitterli2020spatiotemporal}
\bibfield{author}{\bibinfo{person}{Benedikt Bitterli}, \bibinfo{person}{Chris
  Wyman}, \bibinfo{person}{Matt Pharr}, \bibinfo{person}{Peter Shirley},
  \bibinfo{person}{Aaron Lefohn}, {and} \bibinfo{person}{Wojciech Jarosz}.}
  \bibinfo{year}{2020}\natexlab{}.
\newblock \showarticletitle{Spatiotemporal reservoir resampling for real-time
  ray tracing with dynamic direct lighting}.
\newblock \bibinfo{journal}{\emph{ACM Transactions on Graphics (TOG)}}
  \bibinfo{volume}{39}, \bibinfo{number}{4} (\bibinfo{year}{2020}),
  \bibinfo{pages}{148--1}.
\newblock


\bibitem[Dachsbacher et~al\mbox{.}(2007)]%
        {dachsbacher2007implicit}
\bibfield{author}{\bibinfo{person}{Carsten Dachsbacher}, \bibinfo{person}{Marc
  Stamminger}, \bibinfo{person}{George Drettakis}, {and}
  \bibinfo{person}{Fr{\'e}do Durand}.} \bibinfo{year}{2007}\natexlab{}.
\newblock \showarticletitle{Implicit visibility and antiradiance for
  interactive global illumination}.
\newblock \bibinfo{journal}{\emph{ACM Transactions on Graphics (TOG)}}
  \bibinfo{volume}{26}, \bibinfo{number}{3} (\bibinfo{year}{2007}),
  \bibinfo{pages}{61--es}.
\newblock


\bibitem[Diolatzis et~al\mbox{.}(2022)]%
        {diolatzis2022active}
\bibfield{author}{\bibinfo{person}{Stavros Diolatzis}, \bibinfo{person}{Julien
  Philip}, {and} \bibinfo{person}{George Drettakis}.}
  \bibinfo{year}{2022}\natexlab{}.
\newblock \showarticletitle{Active Exploration for Neural Global Illumination
  of Variable Scenes}.
\newblock \bibinfo{journal}{\emph{ACM Transactions on Graphics}}
  (\bibinfo{year}{2022}).
\newblock


\bibitem[Gharbi et~al\mbox{.}(2019)]%
        {gharbi2019sample}
\bibfield{author}{\bibinfo{person}{Micha{\"e}l Gharbi},
  \bibinfo{person}{Tzu-Mao Li}, \bibinfo{person}{Miika Aittala},
  \bibinfo{person}{Jaakko Lehtinen}, {and} \bibinfo{person}{Fr{\'e}do Durand}.}
  \bibinfo{year}{2019}\natexlab{}.
\newblock \showarticletitle{Sample-based Monte Carlo denoising using a
  kernel-splatting network}.
\newblock \bibinfo{journal}{\emph{ACM Transactions on Graphics (TOG)}}
  \bibinfo{volume}{38}, \bibinfo{number}{4} (\bibinfo{year}{2019}),
  \bibinfo{pages}{1--12}.
\newblock


\bibitem[Granskog et~al\mbox{.}(2020)]%
        {granskog2020compositional}
\bibfield{author}{\bibinfo{person}{Jonathan Granskog}, \bibinfo{person}{Fabrice
  Rousselle}, \bibinfo{person}{Marios Papas}, {and} \bibinfo{person}{Jan
  Nov{\'a}k}.} \bibinfo{year}{2020}\natexlab{}.
\newblock \showarticletitle{Compositional neural scene representations for
  shading inference}.
\newblock \bibinfo{journal}{\emph{ACM Transactions on Graphics (TOG)}}
  \bibinfo{volume}{39}, \bibinfo{number}{4} (\bibinfo{year}{2020}),
  \bibinfo{pages}{135--1}.
\newblock


\bibitem[Hasselgren et~al\mbox{.}(2020)]%
        {hasselgren2020neural}
\bibfield{author}{\bibinfo{person}{Jon Hasselgren}, \bibinfo{person}{Jacob
  Munkberg}, \bibinfo{person}{Marco Salvi}, \bibinfo{person}{Anjul Patney},
  {and} \bibinfo{person}{Aaron Lefohn}.} \bibinfo{year}{2020}\natexlab{}.
\newblock \showarticletitle{Neural temporal adaptive sampling and denoising}.
  In \bibinfo{booktitle}{\emph{Computer Graphics Forum}},
  Vol.~\bibinfo{volume}{39}. Wiley Online Library, \bibinfo{pages}{147--155}.
\newblock


\bibitem[I\c{s}{\i}k et~al\mbox{.}(2021)]%
        {isik2021anf}
\bibfield{author}{\bibinfo{person}{Mustafa I\c{s}{\i}k},
  \bibinfo{person}{Krishna Mullia}, \bibinfo{person}{Matthew Fisher},
  \bibinfo{person}{Jonathan Eisenmann}, {and} \bibinfo{person}{Micha\"{e}l
  Gharbi}.} \bibinfo{year}{2021}\natexlab{}.
\newblock \showarticletitle{Interactive Monte Carlo Denoising using Affinity of
  Neural Features}.
\newblock \bibinfo{journal}{\emph{ACM Transactions on Graphics (TOG)}}
  \bibinfo{volume}{40}, \bibinfo{number}{4}, Article \bibinfo{articleno}{37}
  (\bibinfo{year}{2021}), \bibinfo{numpages}{13}~pages.
\newblock
\urldef\tempurl%
\url{https://doi.org/10.1145/3450626.3459793}
\showDOI{\tempurl}


\bibitem[Kelemen et~al\mbox{.}(2002)]%
        {kelemen2002simple}
\bibfield{author}{\bibinfo{person}{Csaba Kelemen},
  \bibinfo{person}{L{\'a}szl{\'o} Szirmay-Kalos}, \bibinfo{person}{Gy{\"o}rgy
  Antal}, {and} \bibinfo{person}{Ferenc Csonka}.}
  \bibinfo{year}{2002}\natexlab{}.
\newblock \showarticletitle{A simple and robust mutation strategy for the
  metropolis light transport algorithm}. In \bibinfo{booktitle}{\emph{Computer
  Graphics Forum}}, Vol.~\bibinfo{volume}{21}. Wiley Online Library,
  \bibinfo{pages}{531--540}.
\newblock


\bibitem[Keller et~al\mbox{.}(2015)]%
        {keller2015path}
\bibfield{author}{\bibinfo{person}{Alexander Keller}, \bibinfo{person}{Luca
  Fascione}, \bibinfo{person}{Marcos Fajardo}, \bibinfo{person}{Iliyan
  Georgiev}, \bibinfo{person}{Per Christensen}, \bibinfo{person}{Johannes
  Hanika}, \bibinfo{person}{Christian Eisenacher}, {and}
  \bibinfo{person}{Gregory Nichols}.} \bibinfo{year}{2015}\natexlab{}.
\newblock \showarticletitle{The path tracing revolution in the movie industry}.
\newblock In \bibinfo{booktitle}{\emph{ACM SIGGRAPH 2015 Courses}}.
  \bibinfo{pages}{1--7}.
\newblock


\bibitem[Kriv{\'a}nek et~al\mbox{.}(2005)]%
        {krivanek2005radiance}
\bibfield{author}{\bibinfo{person}{Jaroslav Kriv{\'a}nek},
  \bibinfo{person}{Pascal Gautron}, \bibinfo{person}{Sumanta Pattanaik}, {and}
  \bibinfo{person}{Kadi Bouatouch}.} \bibinfo{year}{2005}\natexlab{}.
\newblock \showarticletitle{Radiance caching for efficient global illumination
  computation}.
\newblock \bibinfo{journal}{\emph{IEEE Transactions on Visualization and
  Computer Graphics}} \bibinfo{volume}{11}, \bibinfo{number}{5}
  (\bibinfo{year}{2005}), \bibinfo{pages}{550--561}.
\newblock


\bibitem[Kuznetsov et~al\mbox{.}(2018)]%
        {kuznetsov2018deep}
\bibfield{author}{\bibinfo{person}{Alexandr Kuznetsov},
  \bibinfo{person}{Nima~Khademi Kalantari}, {and} \bibinfo{person}{Ravi
  Ramamoorthi}.} \bibinfo{year}{2018}\natexlab{}.
\newblock \showarticletitle{Deep Adaptive Sampling for Low Sample Count
  Rendering}.
\newblock \bibinfo{journal}{\emph{Computer Graphics Forum}}
  \bibinfo{volume}{37} (\bibinfo{year}{2018}), \bibinfo{pages}{35--44}.
\newblock


\bibitem[Lin et~al\mbox{.}(2022)]%
        {lin2022generalized}
\bibfield{author}{\bibinfo{person}{Daqi Lin}, \bibinfo{person}{Markus
  Kettunen}, \bibinfo{person}{Benedikt Bitterli}, \bibinfo{person}{Jacopo
  Pantaleoni}, \bibinfo{person}{Cem Yuksel}, {and} \bibinfo{person}{Chris
  Wyman}.} \bibinfo{year}{2022}\natexlab{}.
\newblock \showarticletitle{Generalized resampled importance sampling:
  foundations of ReSTIR}.
\newblock \bibinfo{journal}{\emph{ACM Transactions on Graphics (TOG)}}
  \bibinfo{volume}{41}, \bibinfo{number}{4} (\bibinfo{year}{2022}),
  \bibinfo{pages}{1--23}.
\newblock


\bibitem[Loos et~al\mbox{.}(2011)]%
        {loos2011modular}
\bibfield{author}{\bibinfo{person}{Bradford~J Loos}, \bibinfo{person}{Lakulish
  Antani}, \bibinfo{person}{Kenny Mitchell}, \bibinfo{person}{Derek
  Nowrouzezahrai}, \bibinfo{person}{Wojciech Jarosz}, {and}
  \bibinfo{person}{Peter-Pike Sloan}.} \bibinfo{year}{2011}\natexlab{}.
\newblock \showarticletitle{Modular radiance transfer}. In
  \bibinfo{booktitle}{\emph{Proceedings of the 2011 SIGGRAPH Asia Conference}}.
  \bibinfo{pages}{1--10}.
\newblock


\bibitem[Loos et~al\mbox{.}(2012)]%
        {loos2012delta}
\bibfield{author}{\bibinfo{person}{Bradford~J Loos}, \bibinfo{person}{Derek
  Nowrouzezahrai}, \bibinfo{person}{Wojciech Jarosz}, {and}
  \bibinfo{person}{Peter-Pike Sloan}.} \bibinfo{year}{2012}\natexlab{}.
\newblock \showarticletitle{Delta radiance transfer}. In
  \bibinfo{booktitle}{\emph{Proceedings of the ACM SIGGRAPH Symposium on
  Interactive 3D Graphics and Games}}. \bibinfo{pages}{191--196}.
\newblock


\bibitem[M{\"u}ller et~al\mbox{.}(2021)]%
        {NeuralCache}
\bibfield{author}{\bibinfo{person}{Thomas M{\"u}ller}, \bibinfo{person}{Fabrice
  Rousselle}, \bibinfo{person}{Jan Nov{\'a}k}, {and} \bibinfo{person}{Alexander
  Keller}.} \bibinfo{year}{2021}\natexlab{}.
\newblock \showarticletitle{Real-time neural radiance caching for path
  tracing}.
\newblock \bibinfo{journal}{\emph{ACM Transactions on Graphics (TOG)}}
  \bibinfo{volume}{40}, \bibinfo{number}{4} (\bibinfo{year}{2021}),
  \bibinfo{pages}{1--16}.
\newblock


\bibitem[Nalbach et~al\mbox{.}(2017)]%
        {Nalbach2017b}
\bibfield{author}{\bibinfo{person}{Oliver Nalbach}, \bibinfo{person}{Elena
  Arabadzhiyska}, \bibinfo{person}{Dushyant Mehta}, \bibinfo{person}{Hans-Peter
  Seidel}, {and} \bibinfo{person}{Tobias Ritschel}.}
  \bibinfo{year}{2017}\natexlab{}.
\newblock \showarticletitle{Deep Shading: Convolutional Neural Networks for
  Screen-Space Shading}.
\newblock \bibinfo{journal}{\emph{Computer Graphics Forum (Proc. EGSR 2017)}}
  \bibinfo{volume}{36}, \bibinfo{number}{4} (\bibinfo{year}{2017}),
  \bibinfo{pages}{65--78}.
\newblock


\bibitem[Ouyang et~al\mbox{.}(2021)]%
        {ouyang2021restir}
\bibfield{author}{\bibinfo{person}{Yaobin Ouyang}, \bibinfo{person}{Shiqiu
  Liu}, \bibinfo{person}{Markus Kettunen}, \bibinfo{person}{Matt Pharr}, {and}
  \bibinfo{person}{Jacopo Pantaleoni}.} \bibinfo{year}{2021}\natexlab{}.
\newblock \showarticletitle{ReSTIR GI: Path resampling for real-time path
  tracing}. In \bibinfo{booktitle}{\emph{Computer Graphics Forum}},
  Vol.~\bibinfo{volume}{40}. Wiley Online Library, \bibinfo{pages}{17--29}.
\newblock


\bibitem[Ren et~al\mbox{.}(2013)]%
        {ren2013global}
\bibfield{author}{\bibinfo{person}{Peiran Ren}, \bibinfo{person}{Jiaping Wang},
  \bibinfo{person}{Minmin Gong}, \bibinfo{person}{Stephen Lin},
  \bibinfo{person}{Xin Tong}, {and} \bibinfo{person}{Baining Guo}.}
  \bibinfo{year}{2013}\natexlab{}.
\newblock \showarticletitle{Global illumination with radiance regression
  functions}.
\newblock \bibinfo{journal}{\emph{ACM Transactions on Graphics (TOG)}}
  \bibinfo{volume}{32}, \bibinfo{number}{4} (\bibinfo{year}{2013}),
  \bibinfo{pages}{130}.
\newblock


\bibitem[Rousselle et~al\mbox{.}(2016)]%
        {rousselle2016image}
\bibfield{author}{\bibinfo{person}{Fabrice Rousselle},
  \bibinfo{person}{Wojciech Jarosz}, {and} \bibinfo{person}{Jan Nov{\'a}k}.}
  \bibinfo{year}{2016}\natexlab{}.
\newblock \showarticletitle{Image-space control variates for rendering}.
\newblock \bibinfo{journal}{\emph{ACM Transactions on Graphics (TOG)}}
  \bibinfo{volume}{35}, \bibinfo{number}{6} (\bibinfo{year}{2016}),
  \bibinfo{pages}{1--12}.
\newblock


\bibitem[Salehi et~al\mbox{.}(2022)]%
        {salehi2022deep}
\bibfield{author}{\bibinfo{person}{Farnood Salehi}, \bibinfo{person}{Marco
  Manzi}, \bibinfo{person}{Gerhard Roethlin}, \bibinfo{person}{Romann Weber},
  \bibinfo{person}{Christopher Schroers}, {and} \bibinfo{person}{Marios
  Papas}.} \bibinfo{year}{2022}\natexlab{}.
\newblock \showarticletitle{Deep Adaptive Sampling and Reconstruction using
  Analytic Distributions}.
\newblock \bibinfo{journal}{\emph{ACM Transactions on Graphics (TOG)}}
  \bibinfo{volume}{41}, \bibinfo{number}{6} (\bibinfo{year}{2022}),
  \bibinfo{pages}{1--16}.
\newblock


\bibitem[Seyb et~al\mbox{.}(2020)]%
        {seyb20uberbake}
\bibfield{author}{\bibinfo{person}{Dario Seyb}, \bibinfo{person}{Peter-Pike
  Sloan}, \bibinfo{person}{Ari Silvennoinen}, \bibinfo{person}{Michał
  Iwanicki}, {and} \bibinfo{person}{Wojciech Jarosz}.}
  \bibinfo{year}{2020}\natexlab{}.
\newblock \showarticletitle{The design and evolution of the {{UberBake}} light
  baking system}.
\newblock \bibinfo{journal}{\emph{ACM Transactions on Graphics (Proceedings of
  SIGGRAPH)}} \bibinfo{volume}{39}, \bibinfo{number}{4} (\bibinfo{date}{July}
  \bibinfo{year}{2020}).
\newblock
\urldef\tempurl%
\url{https://doi.org/10/gg8xc9}
\showDOI{\tempurl}


\bibitem[Silvennoinen and Sloan(2021)]%
        {silvennoinen2021moving}
\bibfield{author}{\bibinfo{person}{Ari Silvennoinen} {and}
  \bibinfo{person}{Peter-Pike Sloan}.} \bibinfo{year}{2021}\natexlab{}.
\newblock \showarticletitle{Moving basis decomposition for precomputed light
  transport}. In \bibinfo{booktitle}{\emph{Computer Graphics Forum}},
  Vol.~\bibinfo{volume}{40}. Wiley Online Library, \bibinfo{pages}{127--137}.
\newblock


\bibitem[Tewari et~al\mbox{.}(2022)]%
        {tewari2022advances}
\bibfield{author}{\bibinfo{person}{Ayush Tewari}, \bibinfo{person}{Justus
  Thies}, \bibinfo{person}{Ben Mildenhall}, \bibinfo{person}{Pratul
  Srinivasan}, \bibinfo{person}{Edgar Tretschk}, \bibinfo{person}{Wang Yifan},
  \bibinfo{person}{Christoph Lassner}, \bibinfo{person}{Vincent Sitzmann},
  \bibinfo{person}{Ricardo Martin-Brualla}, \bibinfo{person}{Stephen Lombardi},
  {et~al\mbox{.}}} \bibinfo{year}{2022}\natexlab{}.
\newblock \showarticletitle{Advances in neural rendering}. In
  \bibinfo{booktitle}{\emph{Computer Graphics Forum}},
  Vol.~\bibinfo{volume}{41}. Wiley Online Library, \bibinfo{pages}{703--735}.
\newblock


\bibitem[Thomas et~al\mbox{.}(2022)]%
        {thomas2022temporally}
\bibfield{author}{\bibinfo{person}{Manu~Mathew Thomas}, \bibinfo{person}{Gabor
  Liktor}, \bibinfo{person}{Christoph Peters}, \bibinfo{person}{Sungye Kim},
  \bibinfo{person}{Karthik Vaidyanathan}, {and} \bibinfo{person}{Angus~G
  Forbes}.} \bibinfo{year}{2022}\natexlab{}.
\newblock \showarticletitle{Temporally Stable Real-Time Joint Neural Denoising
  and Supersampling}.
\newblock \bibinfo{journal}{\emph{Proceedings of the ACM on Computer Graphics
  and Interactive Techniques}} \bibinfo{volume}{5}, \bibinfo{number}{3}
  (\bibinfo{year}{2022}), \bibinfo{pages}{1--22}.
\newblock


\bibitem[Ward et~al\mbox{.}(1988)]%
        {ward1988ray}
\bibfield{author}{\bibinfo{person}{Gregory~J Ward}, \bibinfo{person}{Francis~M
  Rubinstein}, {and} \bibinfo{person}{Robert~D Clear}.}
  \bibinfo{year}{1988}\natexlab{}.
\newblock \showarticletitle{A ray tracing solution for diffuse
  interreflection}. In \bibinfo{booktitle}{\emph{Proceedings of the 15th annual
  conference on Computer graphics and interactive techniques}}.
  \bibinfo{pages}{85--92}.
\newblock


\bibitem[Zheng et~al\mbox{.}(2023)]%
        {zheng2023nelt}
\bibfield{author}{\bibinfo{person}{Chuankun Zheng}, \bibinfo{person}{Yuchi
  Huo}, \bibinfo{person}{Shaohua Mo}, \bibinfo{person}{Zhihua Zhong},
  \bibinfo{person}{Zhizhen Wu}, \bibinfo{person}{Wei Hua}, \bibinfo{person}{Rui
  Wang}, {and} \bibinfo{person}{Hujun Bao}.} \bibinfo{year}{2023}\natexlab{}.
\newblock \showarticletitle{Nelt: Object-oriented neural light transfer}.
\newblock \bibinfo{journal}{\emph{ACM Transactions on Graphics}}
  \bibinfo{volume}{42}, \bibinfo{number}{5} (\bibinfo{year}{2023}),
  \bibinfo{pages}{1--16}.
\newblock


\bibitem[Zhou et~al\mbox{.}(2005)]%
        {zhou2005precomputed}
\bibfield{author}{\bibinfo{person}{Kun Zhou}, \bibinfo{person}{Yaohua Hu},
  \bibinfo{person}{Stephen Lin}, \bibinfo{person}{Baining Guo}, {and}
  \bibinfo{person}{Heung-Yeung Shum}.} \bibinfo{year}{2005}\natexlab{}.
\newblock \showarticletitle{Precomputed shadow fields for dynamic scenes}.
\newblock In \bibinfo{booktitle}{\emph{ACM SIGGRAPH 2005 Papers}}.
  \bibinfo{pages}{1196--1201}.
\newblock


\bibitem[Zwicker et~al\mbox{.}(2015)]%
        {zwicker2015recent}
\bibfield{author}{\bibinfo{person}{Matthias Zwicker}, \bibinfo{person}{Wojciech
  Jarosz}, \bibinfo{person}{Jaakko Lehtinen}, \bibinfo{person}{Bochang Moon},
  \bibinfo{person}{Ravi Ramamoorthi}, \bibinfo{person}{Fabrice Rousselle},
  \bibinfo{person}{Pradeep Sen}, \bibinfo{person}{Cyril Soler}, {and}
  \bibinfo{person}{S-E Yoon}.} \bibinfo{year}{2015}\natexlab{}.
\newblock \showarticletitle{Recent advances in adaptive sampling and
  reconstruction for Monte Carlo rendering}. In
  \bibinfo{booktitle}{\emph{Computer graphics forum}},
  Vol.~\bibinfo{volume}{34}. Wiley Online Library, \bibinfo{pages}{667--681}.
\newblock


\end{thebibliography}

\appendix

\begin{figure*}[]
\centering
\includegraphics[width=.95\linewidth]{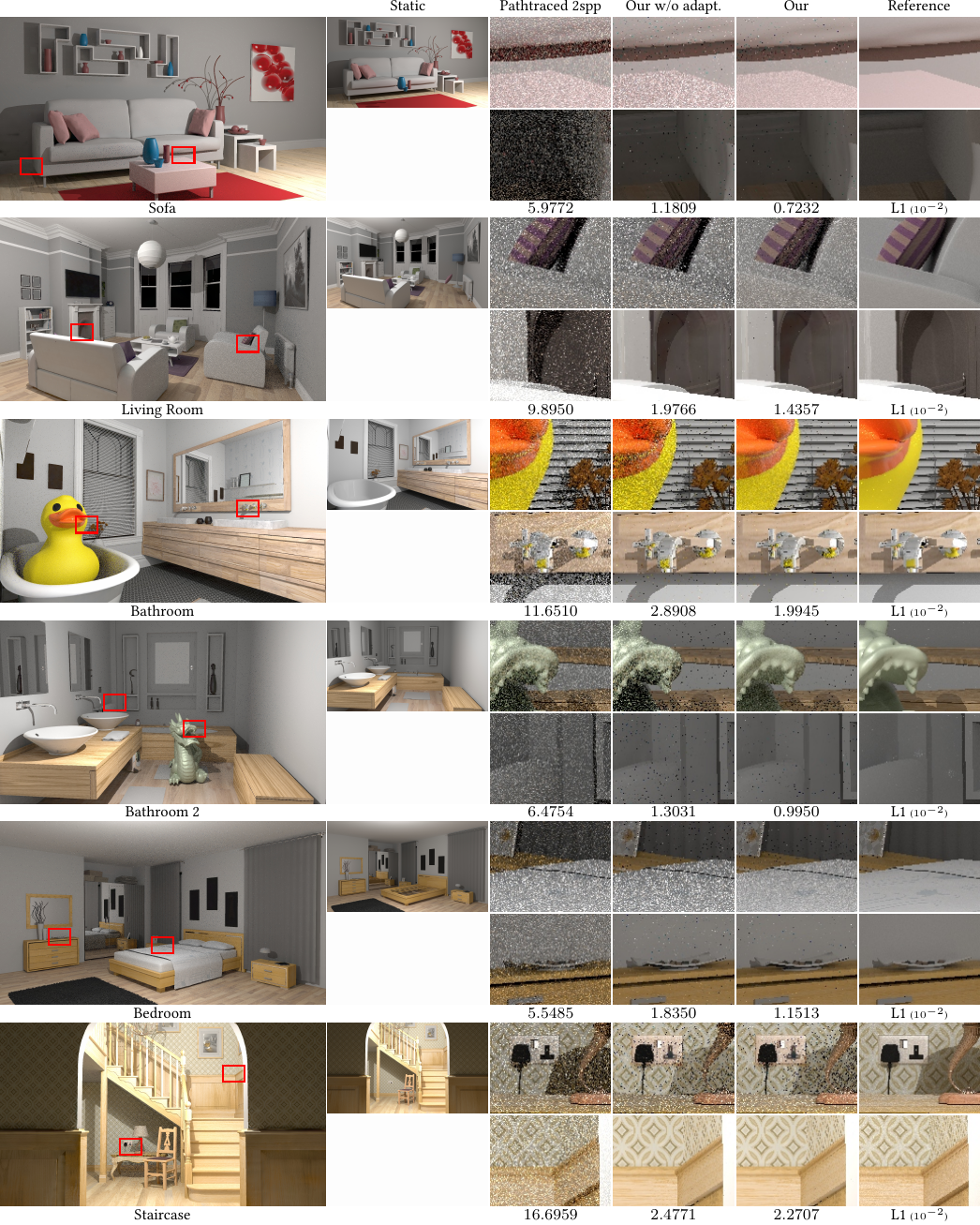}
\caption{
Example of our method on different scenes. For each scene, we added a new dynamic object. The original scenes are labeled as static. The 1st column shows path-traced results with 2spp. The next column shows our results with 1spp. It  already has  3 to 6 times better quality. The second to last column is our method with adaptive sampling. By redistributing samples to the region of change, we achieve up to 8 times improvement.
}
\label{fig:comp1}
\end{figure*}
\begin{figure*}[]
\centering
\includegraphics[width=1\linewidth]{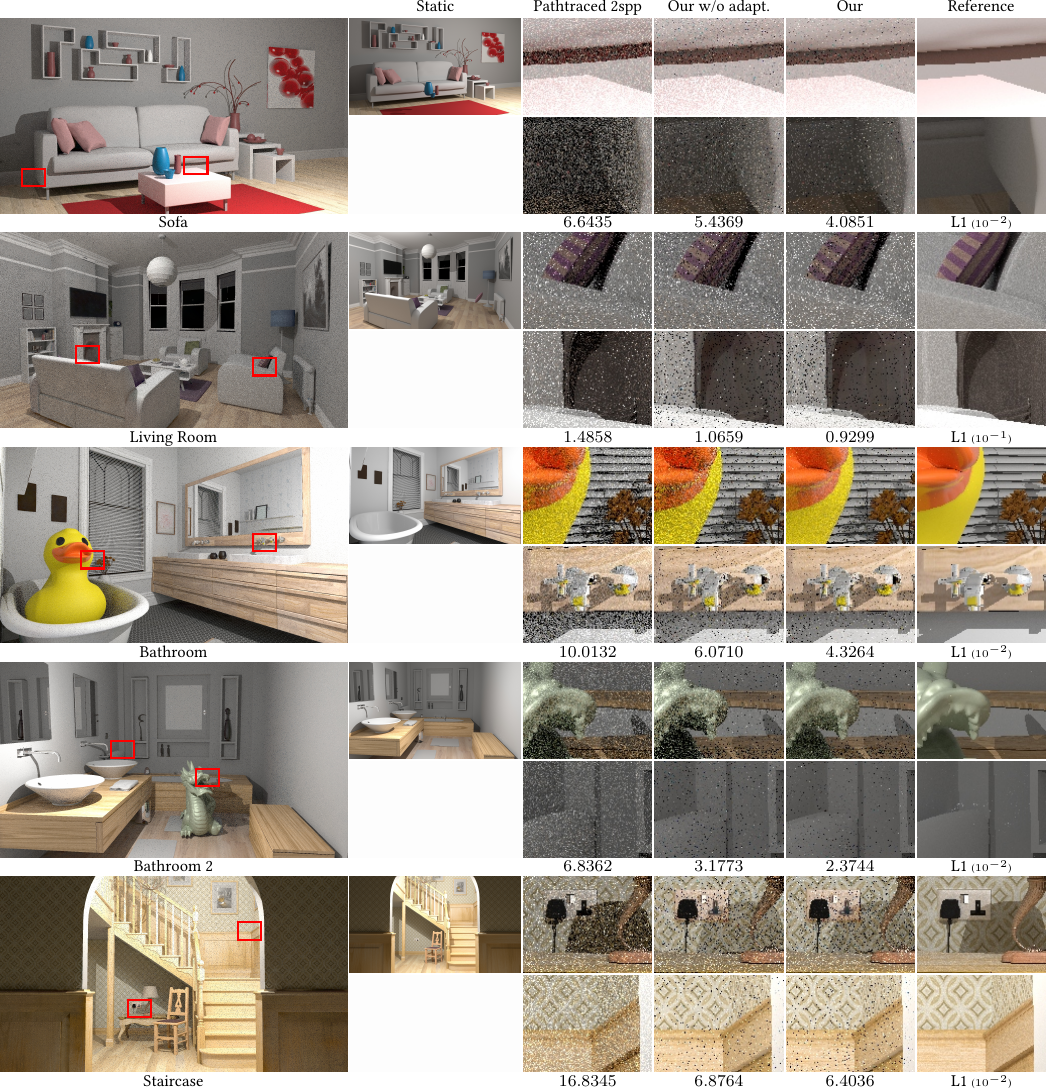}
\caption{
Scenes at different illumination and added objects. When a scene has both kinds of changes, the difference contribution, $\Ldelta$ becomes more intense and therefore noisier. Despite the amount of changes, our method is still able to outperform the path-tracing.
}
\label{fig:comp_lights}
\end{figure*}

\end{document}